\documentclass[conference]{IEEEtran}

\usepackage{booktabs}

\usepackage{makecell}

\usepackage{wrapfig}

\usepackage{mathptmx}       
\usepackage{helvet}         
\usepackage{courier}        
\usepackage{type1cm}        
\usepackage{makeidx}         
\usepackage{graphicx}        
\usepackage{multicol}        
\usepackage[bottom]{footmisc}

\usepackage{subfigure}

\usepackage{cite}

\usepackage[hyphens]{url}
\usepackage{breakurl}

\usepackage{dsfont}

\usepackage{todonotes}
\usepackage{color}

\usepackage{xspace}
\newcommand{\etal}{\textit{et al.}\xspace}

\newcommand{\ie}{\textit{i.e.,}\xspace}

\mathchardef\mhyphen="2D

\renewcommand{\paragraph}[1]{\medskip \noindent \textbf{#1.\ }}

\begin{document}

\title{A Review of Privacy and Consent Management in Healthcare: A Focus on Emerging Data Sources}

\author{
\IEEEauthorblockN{Muhammad Rizwan Asghar}
\IEEEauthorblockA{Cyber Security Foundry\\
The University of Auckland\\
New Zealand}\\
\IEEEauthorblockN{TzeHowe Lee}
\IEEEauthorblockA{Cyber Security Foundry\\
The University of Auckland\\
New Zealand}
\and
\IEEEauthorblockN{Mirza Mansoor Baig}
\IEEEauthorblockA{Research Team\\
Orion Health\\
New Zealand}\\
\IEEEauthorblockN{Ehsan Ullah} 
\IEEEauthorblockA{Clinical Quality and Safety Service\\
Auckland District Health Board\\
New Zealand}
\and
\IEEEauthorblockN{Giovanni Russello}
\IEEEauthorblockA{Cyber Security Foundry\\
The University of Auckland\\
New Zealand} \\
\IEEEauthorblockN{Gillian Dobbie}
\IEEEauthorblockA{Cyber Security Foundry\\
The University of Auckland\\
New Zealand}
}

\maketitle

\begin{abstract}
The emergence of New Data Sources (NDS) in healthcare is revolutionising traditional electronic health records in terms of data availability, storage, and access.
Increasingly, clinicians are using NDS to build a virtual holistic image of a patient’s health condition. 
This research is focused on a review and analysis of the current legislation and privacy rules available for healthcare professionals. 
NDS in this project refers to and includes patient-generated health data, consumer device data, wearable health and fitness data, and data from social media.

This project reviewed legal and regulatory requirements for New Zealand, Australia, the European Union, and the United States to establish the ground reality of existing mechanisms in place concerning the use of NDS. 
The outcome of our research is to recommend changes and enhancements required to better prepare for the 'tsunami' of NDS and applications in the currently evolving data-driven healthcare area and precision or personalised health initiatives such as Precision Driven Health (PDH) in New Zealand.
\end{abstract}

\section{Introduction}
New Data Sources (NDS) include health and medical-related patient information from smart sensors, advanced devices, social media, and genetic data. 
Aggregation of NDS in an Electronic Health Record (EHR) might support and provide precise and tailored healthcare outcomes. 
Although, gathering of multiple data sources (including NDS) enables clinicians to access and view detailed patient information at the point-of-care for precise and informed decision making, it also raises serious privacy concerns, in particular when a patient might not be aware of who is accessing their health data. 
It is predicted that a patient's electronic health records in the near future will have non-traditional health information (including NDS). 
An initiative looking in to the future of healthcare is Precision Driven Health (PDH)\footnote{\url{http://www.precisiondrivenhealth.com}} - a New Zealand healthcare research partnership investing in the future of data-driven healthcare. 

Current systems and processes require an effective way of getting informed consent not only for EHRs but also for the NDS. 
From a privacy point of view, there are two main concerns. 
First, how to capture patient’s informed consent for collecting NDS data. 
Second, how to obtain consent for regulating access to this collected data. This is technically known as access control. For example, a patient might want to allow the General Practitioner (GP) access to the data, but could decide not to provide access to any other healthcare professionals. 
In both cases, the challenging part is to get and manage consent in a flexible, usable, and transparent manner.

The main problem is that consent management for NDS is fragile in nature, \ie a patient can withdraw his/her consent to provide access to the NDS at any time. 
This is different from giving consent for providing access to an EHR that might be necessary for diagnosing health issues. 
Unfortunately, current healthcare systems are not capable of integrating NDS for a number of reasons including technical issues, legal and regulatory requirements, and usability aspects, where patients should be kept in the loop but without too much intervention. 

In \cite{Powles-2017-Google}, Powles and Hodson discuss the case of the data sharing agreement between DeepMind and the Royal Free London NHS Foundation Trust.
They point out unlawful transfer of patient data to third parties. 
The UK law requires ``explicit consent from a patient when identifiable data is passed to a third party, when the third party is not in a direct care relationship with the patient''. 
However, there was no process of obtaining patient consent or giving notices to patients whose data were used. 
There is also concern that the data being used is broader than the original stated purpose of developing an app for Acute Kidney Injury (AKI). 
There is also no justification for the breadth of data transfer and the length of data retention.
The main issue is that DeepMind is not open for public audit and scrutiny but rather is self-maintained by Google.
This case strongly advocates that there is a need to come up with a transparent framework for obtaining and managing consent, where patients are empowered to provide access to their health information and view who is using their health information.

The rest of this paper is organised as follows.
In Section \ref{sec:legal}, we cover legal and regulatory requirements concerning privacy and consent in healthcare.
In Section \ref{sec:approaches}, we review existing research approaches for capturing and managing consent.
In Section \ref{sec:challenges}, we discuss opportunities and challenges. 
Finally, we conclude the paper in Section \ref{sec:conclusion}.

\begin{table*}[htbp]
  \centering
  \caption{Summary of legislation of countries studied and privacy concerns in EHR}
    \label{tab:legislation}%
    \begin{tabular}{|p{7.285em}|p{12em}|p{12.285em}|p{12.355em}|p{13.57em}|}
     \hline
     
    \textbf{\makecell[l]{Legislation/ \\ Concerns}} & \textbf{NZ HPIC 1994} & \textbf{AUS NSW HRIPA} & \textbf{EU DPD} & \textbf{US HIPAA} \\ \hline
    
    \textit{Data collection \& Patient rights} & 
    Rules 1-4 summarise purpose and source, informing individuals of data use and how data is collected. & 
    Individuals must be informed of the purpose and extent of data collected. & 
    Detailed information of how data is being used must be made available when identifiable information is used. & 
    Obligation to inform patients how information is disclosed and to whom in privacy notices. \\ \hline
    
    \textit{Referrals /data sharing} & 
    Rules 10-12 limit the use and disclosure of health information to the stated purposes. 
    Unique identifiers must be used to protect personal information. &
    Health records can only be used for the purposes stated to the patient, any secondary use must be requested unless it is an emergency. & 
    There are no specific rules barring or requesting patient consent before sharing data, although the purpose for each collection must be stated explicitly. & 
    Patient has to specifically request to restrict EHR, but healthcare provider has no obligation to comply. 
    No patient authorisation needed to share data if it is in regards to treatment. \\ \hline
    
    \textit{Ability to view/correct EHR} & 
    Rules 6-8 state the need to give patient rights to view EHR and correct if necessary, also data integrity and retention.
    Health data must be stored for 10 years & 
    Patients can view their EHR and request to delete, change, or add data. & 
    Patients are given rights to view, erase, and correct their EHR data. & 
    Patients are given the right to view and request for corrections but healthcare providers do not have to conform. \\ \hline
    
    \textit{Data disclosure} & 
    Rules 5 and 11 cover disclosure of confidential data to authorised individuals only and for consented purposes. & 
    Disclosure of data is prohibited outside of consented purpose and authorised individual.  & Authentication mechanisms, electronic method of identification and audit logs are required. & 
    Access safeguards and breach notification specifications are addressed. 
    Audit logs must be stored for 6 years. \\ \hline
    
    \end{tabular}%
\end{table*}%

\section{Legal and Regulatory Requirements}
\label{sec:legal}
This research reviews legal and regulatory requirements of countries that have an established electronic patient record in legislation and regulations in healthcare privacy, namely New Zealand (NZ), Australia (AU), the European Union (EU), and the United States (US). 
We identify the differences and similarities in legal requirements between these countries in terms of consent and healthcare privacy. 

In \textbf{New Zealand}, there is a Health Information Privacy Code (HPIC) 1994 \cite{NZ-HPIC}, which is based on the following rules:
\begin{itemize}
 
\item Rules 1-4: The purpose of collection must be specified while collecting patient consent. 
Patients must be adequately informed about what information is being collected and the collection of information must be done in a lawful and fair manner.

\item Rule 5: covers security and storage of health information, where healthcare providers must ensure adequate protection against misuse, loss, and unauthorised access.

\item Rules 6-7: These rules state patient rights to request and gain access to their health information and enable them to request corrections.

\item Rule 8: It ensures that healthcare providers must ensure the information is up-to-date and relevant to the case.

\item Rule 9: It explains retention of healthcare information, where healthcare providers are required to retain information for a minimum of 10 years.

\item Rule 10: It requires the healthcare provider to only use the information for the purposes stated unless permitted by the patient, taken from a public source, or the information has been de-identified.

\item Rule 11: It limits healthcare providers from sharing information.
It explains circumstances under which healthcare providers are allowed to disclose patient information, such as when patients are unable to exercise their rights or someone is acting on their behalf.

\item Rule 12: It governs the use of unique identifiers assigned to patients.
These unique identifiers allow the patient to be anonymous.

\end{itemize}

The regulations in \textbf{Australia} are unique, where each state has their own set of laws and regulations concerning privacy~\cite{okeefe_privacy_2010}. 
We reviewed the New South Wales Health Records and Information Privacy Act 2002 (NSW HRIPA) \cite{NSW_HRIPA} as it is the most recent legislation passed in Australia that relates to health information privacy. 
The basic HRIPA principles are similar to the NZ HPIC, which governs data collection, storage and security, data access and accuracy, limitation of use and data disclosure, and identifiers and anonymity.

Although the \textbf{European Union} legislation has no specific clause that covers privacy of EHR, the Data Protection Directive (DPD) 95/46/EC \cite{DPD-EU} is the legal EU framework that protects personal information including the EHR. 
The core of the data collection is that the data must be “adequate, relevant and not excessive in relation to the purposes for which they are processed”. The purpose of the data collection must also be determined and further collection, if not specified in the original purpose, is not allowed.
Under Article 2(h) of the DPD \cite{DPD-EU}, consent is defined as ``must be freely given, specific and informed''.

HIPAA (Health Insurance Portability and Accountability Act) \cite{HIPAA-US} governs with the \textbf{US} legislation for healthcare privacy by defining rules to ensure that personal health information that is stored by healthcare providers would not be disclosed or used in a way that would violate patient privacy. 
HIPAA covers all individually identifiable health information in any form or media and this is called Protected Health Information (ProHI) \cite{HIPAA-US}. 
One of the central aspects of HIPAA’s Privacy Act is minimum necessary disclosure and access for the intended purpose of the request.
HIPAA’s Privacy Act is one of the more extensive laws that cover different entities requiring policy, personnel, technical, administrative, and physical safeguards to prevent unintended or intended disclosure of data unless permitted or required.
Individuals have general important rights under HIPAA, which include the right to request a copy of the ProHI that the covered entities hold, make amendments to the record, permit or deny use of their ProHI, and also an audit log of their ProHI disclosure. 
Individuals can also request to restrict use or disclosure of their ProHI for reasons of treatment, payment or healthcare; however, entities do not need a patient’s consent to share information when it is used for treatment, payment, or healthcare services.

Table \ref{tab:legislation} summarises healthcare legislation of countries inclued in this study. 
It is important to highlight that none of the countries studied include a provision in legislation for NDS.

\section{Existing Research Approaches}
\label{sec:approaches}
The core security service that all EHR systems try to offer is Access Control for enabling access to only legitimate individuals, controlled and determined by the patient.
To enable access control in EHR databases, there are several proposals based on the Role-Based Access Control (RBAC) model, described in \cite{sandhu_role-based_1996, ferraiolo2001proposed}. 
RBAC models require a system administrator to create roles based on job functions and to assign those roles to individuals. 

One of the problems with RBAC is handling of a team in a multi-team environment, where roles might change and are more dynamic, as is the case in healthcare scenarios where different levels of access are required across a multidisciplinary team. In Singapore, an implied consent model is implemented, giving any member of the health organisation unrestricted access to a patient's medical data. 
Audit logs could deter unauthorised access but they are unavailable to the patient. 
With regards to consent revocation, Singapore assumes an implied consent model because ``explicit consent is practically difficult, especially under emergency circumstances'' \cite{sinha2012electronic}. 
Although there is legislation that is supposed to assure the protection of personal data, authorised personnel are generalised, allowing access to one who is associated with a healthcare provider.

Other healthcare implementations, for example Summary Care Record (SCR) in the United Kingdom (UK), use a smart card based system.
The card could only be used at terminals with card readers and by users with a valid login ID and password \cite{NHS-HealthRec}. 
It is assumed a third party (\ie Registration Authority) assigns roles to users and controls the keys that are assigned to the users \cite{PSNC-Smartcard-model}.

Smart card based solutions have been proposed in the literature. 
Neubauer and  Heurix \cite{heurix2011privacy} proposed the use of microcontroller smart cards that can store user credentials, thus acting as a trusted platform. 
User data is encrypted in the database with the keys that are stored on the card.

Lee and Lee \cite{lee2008cryptographic} also proposed a solution to use keys stored in the smart card, a patient-specific key, and also digital signatures to authenticate parties while maintaining the confidentiality of the data. 
Although such solutions manage cryptographic keys nicely, they do not consider problems such as losing the card or damage to the chip. These problems might make the database inaccessible unless there is an off the band backup of keys.

Hembroff and Muftic~\cite{hembroff2010samson} used a system called SAMSON (Secure Access for Medical Smart Cards Over Networks) that uses smart cards and a combination of PIN and patient's biometric fingerprint to access the EHR. 
It considers emergency features in which the patient is unable to give consent, by giving physicians a special privilege to ``break the glass'' and access the EHR. 
The main drawback of this approach is that it assumes a patient’s smart card is present and would serve as a base for the 3-factor authentication proposed. 

Hu \etal \cite{hu2010hybrid} proposed a Hybrid Public Key Infrastructure (HPKI) that uses smart cards given to users with their public and private keys already assigned. 
There are individual level certificates and organisational level certificates available to healthcare providers. 
Such a solution would be interesting if keys could be revoked (and new cards and keys could be issued) in cases where the cards are damaged or lost. 

There are extensions of public key cryptography to ensure fine-grained access control, such as Key Policy Attribute-Based Encryption (KP-ABE) proposed by Goyal \etal \cite{goyal2006attribute}, where encryption keys are distributed based on attributes assigned. 
Ciphertext Policy Attribute-Based Encryption (CP-ABE) \cite{bethencourt2007ciphertext} tries to improve on KP-ABE, reducing the need to trust the Trusted Authority (TA) to distribute the key correctly by attribute. 
CP-ABE embeds policies into the ciphertext to determine access. 
One of the problems that arises from KP-ABE and CP-ABE is a lack of support for attribute revocation, meaning there is no scalable solution to revoke revoked users from accessing the existing data.  

Wu \etal \cite{wu2013trusted} propose a private key generator, which in this case is a trusted third party \ie a cloud infrastructure.
The proposed idea uses symmetric key cryptography for providing fine-grained access control to specific data to a number of users.
However, this framework does not consider consent revocation. 

Zhang \etal \cite{zhang2016consent} presented Consent Based Access Control (CBAC).
To enable access to an EHR, a consent token will be sent to the server and then the data would be re-encrypted with a new key and sent to the requester. 
Consent tokens have a lifetime and can be revoked by sending a revocation notification, where the server would store it in a Consent Revocation List (CRL). 
However, the proposed system does not discuss which piece of data could be shared or how requesting parties could be authenticated.

Ma and Sartipi \cite{ma2014agent} proposed a framework for sharing medical images from labs using OpenID and OAuth as an authentication and authorisation agent. 
Patient consent is stored offline in a repository and would later be captured and executed by an agent. 
The authors defined a comprehensive e-consent UML class diagram to model a privacy based access control policy but it does not consider an automated way to capture consent. 

Weitzel \etal \cite{weitzel2010web} analysed the feasibility of using REST, OAuth, and OpenSocial to integrate social data into the EHR in the form of a gadget. 
The author considers the possibility of integrating a team-based environment as a social network defined using OpenSocial. 
A defined protocol using REST could use the gadget/dashboard to evaluate the patient condition and next treatment.

Wuyts \etal \cite{wuyts2013integrating} proposed a system that uses XACML (eXtensible Access Control Markup Language), which is based on XML, to define access control. 
The XACML authorisation service is then integrated into a Cross Enterprise Document Sharing (XDS) reference architecture. 
XACML enables fine-grained access control using policies that are a defined set of rules restricting access.

Bahga and Madisetti \cite{bahga2013cloud} proposed Cloud Health Information Systems Technology Architecture (CHISTAR), which supports OAuth to ensure that patients are able to give consent. 
The database is encrypted with a master key and then a key encrypting key is passed to the requester and rotated after a certain number of transactions. 
CHISTAR uses RBAC, which is controlled by an administrator to assign access policies and user roles. 
CHISTAR also considers encrypting the data stored on the server using AES-256 encryption. 

Asghar and Russello \cite{asghar2012flexible} reported a consent policy where a patient is able to communicate their permission to grant access to their EHR. 
They propose a consent evaluator that evaluates consent policies and returns the consent response to the requester. 
They define two types of policies: open and complex. 
Open policies are more general and divided into two types, blacklist and whitelist. 
Complex policies specify conditional expressions that must be satisfied. 
In this system, both the data subject and the data controller can define the access control policies. 

The existing research on access control generally lacks an ability to empower patients to decide who can gain access to their EHR. 
In addition, existing systems do not consider consent revocation as an important functionality to be included in their models. 
Consent for NDS is also another key point that is not considered. 
A solution to authenticate, authorise and ensure that patients have control over their EHR and NDS would be vital to ensuring the highest safety and transparency can be achieved.

\section{Opportunities and Challenges}
\label{sec:challenges}
\subsection{Opportunities}
Currently, there is a gap between the legislation and systems that have been implemented across the world. 
Privacy requirements that are laid out in legislation would empower patients by providing them control over their EHR. 
However, in most of the electronic systems studied, emerging sources are not mentioned and patients have no control or rights other than the ability to request a copy to view and/or edit their partial EHR information.

It is evident that better and timely treatments and medical decisions could be achieved if information from multiple data sources is available to the clinician at the point-of-care.
For instance, a GP can make better prescriptions if provided with regional health information, such as viral infections or other diseases.
Using emerging data sources is also useful in cases where clinicians treat patients who are given consent to access a medical database. 
Harman \cite{harman2014quantifying} used Natural Language Processing (NLP) to analyse and detect symptoms of Post-Traumatic Stress Disorder (PTSD), depression, bipolar disorder, and seasonal affective disorder by analysing Twitter posts. 
In summary, the potential benefits of NDS for research and exploiting them to determine the cause of a disease would offer a breakthrough in the modern big data society.

\subsection{Challenges}
Some of the challenges identified, which require justification/answers when trying to achieve transparency by incorporating NDS with EHRs, include:
\begin{itemize}

\item \emph{Adaptation of new systems with existing healthcare infrastructure.} 
Existing healthcare infrastructure lacks the ability to process and pass on context that could be used to mine  'Big Data'. 
Current healthcare practice lacks a system that is able to collect specific information that is required and request it from the user in an easy way.

\item \emph{Communication and exchange of information.} This could be the main challenge in the world of digital records, where exchanging information across different platforms is complex for various reasons, such as the difference of standards and protocols.
The existing information that is already being stored at geographically dispersed locations and across different organisations must be taken into account and tight integration is required.

\item \emph{Other stakeholders in healthcare.}
Our focus was to review privacy only at the healthcare provider level and we include limited healthcare professionals as user such as GPs and hospital Doctors and Nurses. 
But in practice, EHR could also be handled by other stakeholders, such as insurance companies, wider healthcare professionals, government organisations, and professional governing bodies. 
How to incorporate the parties mentioned without over burdening patients must be thoroughly investigated. 

\item \emph{Data use and storage.} 
One of the major concerns in terms of privacy is the storage and use of data. 
Considering the importance of data, and keeping an accurate patient history, the debate of who should be storing the data is also of concern. 
Both New Zealand Health Information Privacy Code 1994 \cite{NZ-HPIC} and Privacy Act 1993 \cite{1993-privacy} stress that (an agency) ``shall not keep information for longer than is required for the purposes for which the information may be lawfully used''.
The principle mentions that the information can only be used for a specified lawful period of time; however, it does not mention who will decide on this length of time and how to determine the appropriate length. 
When including NDS data, which can be private in nature, the data retention period is a concern from a privacy point of view.

\item \emph{Capturing Consent.}
In this project, we highlight the importance of consent that should be captured in a usable, flexible, and transparent manner, thus empowering patients to have control over their data as well as helping clinicians to make better decisions. 
Consent must be simple and presented in a way such that patients should be able to understand and respond appropriately. 
It should be customisable and adaptable to suit the patient’s privacy requirements and healthcare needs. 
Transparency can help answer the questions who, where, what, when, and why the data is collected and used. Usage should be reported back to the patient (if the information is used out-of-scope).

\end{itemize}

\section{Conclusions and Future Work}
\label{sec:conclusion}
New Zealand’s current EHR landscape is fragmented and lacks a centralised collection of patient data and electronic consent management processes and protocols.
In order to incorporate the emerging New Data Sources (NDS), one challenge that should be tackled is how patient data can be unified. 
As our study of processes and protocols highlighted, NDS is not considered in the current legislation and access control to health-related information is not always determined by the patient. 
The main concern is that consent for NDS is fragile in nature and current electronic systems are not capable of handling consents over NDS appropriately. For instance, a patient can freely deny access to their NDS, but healthcare providers/professionals can still access a patient's EHR due to the lack of the inability to combine consent of EHR and NDS into one. 
A dynamic consent is needed where patients can freely approve or withdraw their consent at any time, while staying informed of how their data is being used and by whom.

\section*{Acknowledgment}
This research was supported by funding from the Precision Driven Health research partnership.

\bibliographystyle{IEEEtran}


\end{document}